\documentclass[12pt]{article}
\usepackage{graphicx}
\usepackage{scicite}
\usepackage{times}
\newenvironment{sciabstract}{
\begin{quote} \bf}
{\end{quote}}

\newcounter{lastnote}

\title{Chiral anomaly without relativity}
\author{A.A. Burkov \\ 
\\ \normalsize{Department of Physics and Astronomy, University of Waterloo, Waterloo, Ontario 
N2L 3G1, Canada, and} \\
\normalsize{ITMO University, Saint Petersburg 197101, Russia}}
\date{}
\begin{document}
\baselineskip24pt
\maketitle 
\begin{sciabstract}
Perspective on J. Xiong {\em et al.}, Science {\bf 350}, 413 (2015).  
\end{sciabstract}
The Dirac equation, which describes relativistic fermions, has a mathematically inevitable, but puzzling feature: negative energy solutions. 
The physical reality of these solutions is unquestionable, as one of their direct consequences, the existence of antimatter, is confirmed by experiment. 
It is their interpretation that has always been somewhat controversial. Dirac's own idea was to view the 
vacuum as a state in which all the negative energy levels are physically filled by fermions, which is now known as the Dirac sea. 
This idea seems to directly contradict a common-sense view of the vacuum as a state in which matter is absent and is thus generally disliked among high-energy physicists, 
who prefer to regard the Dirac sea as not much more than a useful metaphor.
On the other hand, the Dirac sea is a very natural concept from the point of view of a condensed matter physicist, since there is a direct and simple analogy: filled valence bands 
of an insulating crystal. 
There exists, however, a phenomenon within the context of the relativistic quantum field theory itself, whose satisfactory understanding seems to be hard to achieve without assigning physical reality to the Dirac sea. 
This phenomenon is the chiral anomaly, a quantum-mechanical violation of chiral symmetry, which was first observed experimentally in the particle physics setting as a decay of a neutral pion into two photons. 
Now, on page 413 of this issue, Xiong {\em et al.} report the first unambiguous observation of this phenomenon in a condensed matter system, a crystal of Na$_3$Bi, manifesting 
as an unusual {\em negative longitudinal magnetoresistance}. This observation makes the vacuum to insulating crystal analogy all the more tangible. 

The chiral anomaly is an unexpected and at first sight even mysterious feature of the relativistic quantum field theory. 
If one applies the Dirac equation to a hypothetical massless fermion (in fact, all observed fermion masses are negligibly small compared to the natural mass scale, the Planck mass), 
one finds that massless fermions possess a strictly conserved physical quantity called chirality, left- or right-handed, which refers to the handedness of their internal angular 
momentum (called spin) relative to the direction of their linear momentum. 
This conservation of chirality may be viewed as a consequence of {\em chiral symmetry} of the Dirac equation for massless particles: it has no preference for either chirality and does not  
mix the two chiralities. 
However, when one passes from the Dirac equation to the corresponding relativistic field theory (which is made unavoidable by the negative-energy solutions of the Dirac equation), 
one finds that the seemingly obvious chiral symmetry disappears, once account is taken of another fundamental physical principle, that of gauge invariance. 
Chirality is no longer conserved when the fermions are placed in an electromagnetic field with collinear electric and magnetic components. 
This was first independently discovered by Adler and by Bell and Jackiw~\cite{Adler69,Jackiw69}, who were trying to explain the observed decay of a neutral pion into two photons, which seemed to be prohibited by the chiral symmetry. 

The chiral anomaly is in fact not that mysterious and is easy to understand physically, but only so when one takes the view that the Dirac sea is real~\cite{Nielsen83}.
In this case, it follows almost immediately from the form of the energy eigenvalues of the Dirac equation for a massless left- or right-handed charged fermion, in the presence of a constant magnetic field. 
As illustrated in the figure, these have the form of the so called Landau levels, discrete energy levels, which disperse continuously as a function of the component of the linear momentum along the direction of the field. 
Handedness of the particles is reflected in the existence in each case of a special Landau level, whose dispersion is chiral, i.e. has a slope of a specific sign, positive or negative. 
This means, in particular, that, while all other Landau levels have either strictly positive or strictly negative energy, the chiral Landau levels necessarily contain both negative and positive 
energy states. 
Chiral anomaly arises as a direct consequence of the existence of these chiral Landau levels. 
Invoking the Dirac sea picture, all the negative energy states are filled by fermions, while all the positive energy states are empty. 
Now suppose that in addition to the magnetic field, an electric field is applied, in the same direction. 
The electric field will accelerate the particles, which means their momentum will change with time. 
This implies, as illustrated in the figure, simultaneous production of particles of one chirality and antiparticles of the opposite one. 
The total charge is thus conserved, while the chiral one is not. 

\begin{figure}[t]
  \includegraphics[width=13cm]{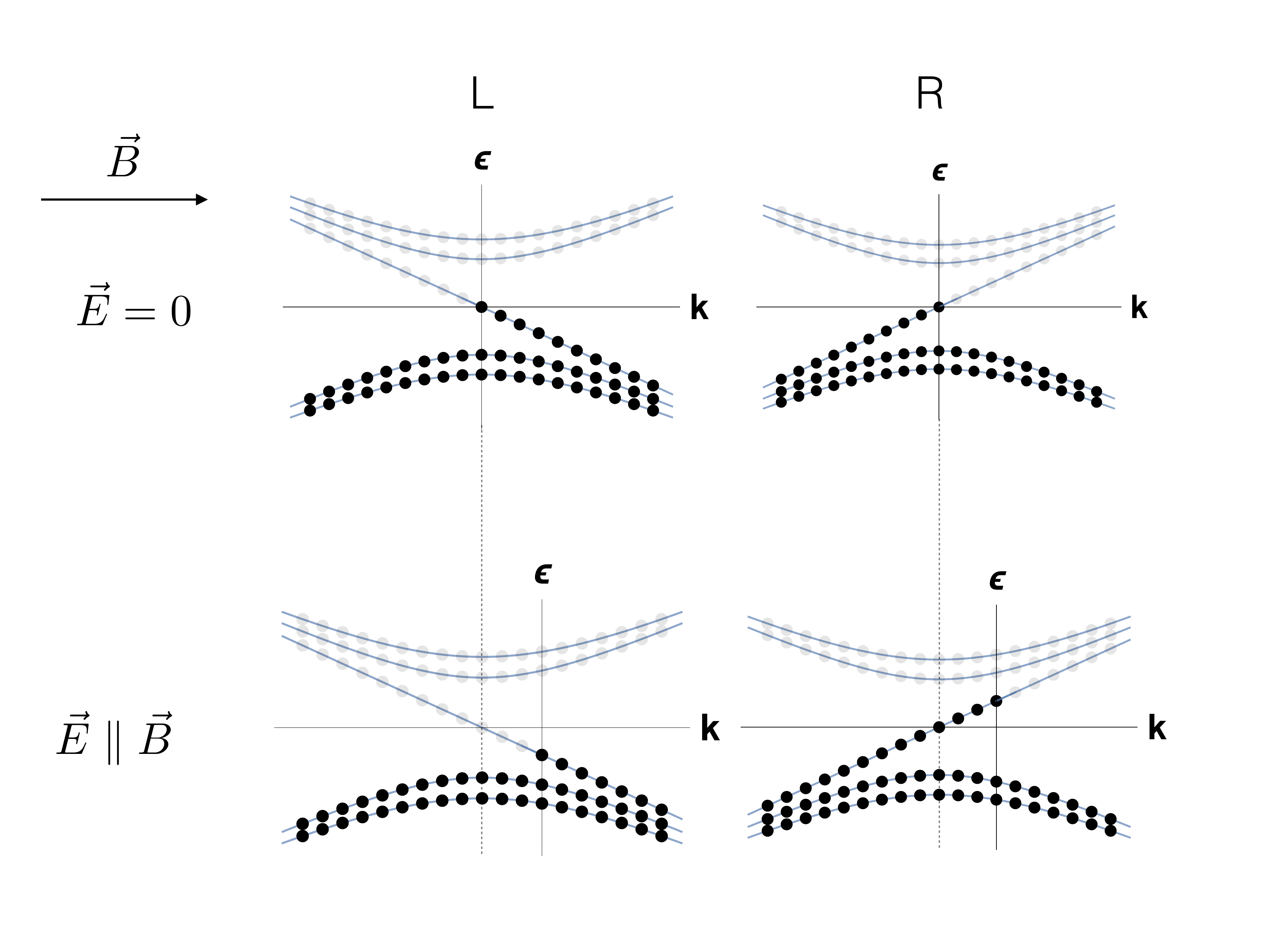}
  \caption{{\bf Illustration of the chiral anomaly}.Top panel: Energy spectrum of the left-handed (L) and the right-handed (R) fermions in the presence of a magnetic field $\vec B$. Filled states with negative energy are shown by black dots, while empty states with positive energy by gray dots. Bottom panel: Same spectrum, but in the presence, in addition, of an electric field $\vec E$, parallel to the magnetic field $\vec B$. Right-handed particles and left-handed antiparticles have been produced.}
  \label{fig:1}
\end{figure}

The observation of this phenomenon in a condensed matter setting, reported by Xiong {\em et al.}, was made possible by an earlier, but also very recent discovery of the Weyl and Dirac semimetals, which are crystalline materials, whose electronic structure mimics the energy-momentum relation of relativistic fermions~\cite{Wan11,Burkov11-1,Shen14,Hasan15-Sci,Lu15-Sci}.
Na$_3$Bi, the specific material studied by Xiong {\em et al.}, is a Dirac semimetal, which means that the left- and right-chirality electrons coexist at the same point in the crystal momentum space. This, however, is not important for the reported phenomena, similar effect should be observed in Weyl semimetals, where opposite-chirality fermions 
exist at distinct points in momentum space. 
The way that the chiral anomaly manifests in Na$_3$Bi is through {\em magnetoresistance}, i.e. a dependence of the electrical resistance of the material on an applied magnetic field.
The physical picture of the chiral anomaly, described in the previous paragraph, when applied to a Dirac or Weyl semimetal, implies a magnetic-field-dependent
contribution to the resistance, which is {\em negative} (i.e. the resistance is reduced and the material becomes a better conductor when the magnetic field is applied) and quadratic in the 
field~\cite{Spivak12,Burkov15-2}. It also exists only when the current is aligned with the direction of the field (i.e. the magnetoresistance is longitudinal), survives up to a very high temperature (about 90 K), and is 
very large (quickly rising to more than 100\% as the temperature decreases below 90 K). 
These features are very unusual and can not be explained by any other known mechanism, but the chiral anomaly. 

What makes the observed effect particularly important, apart from the analogy to particle physics, is that the chiral anomaly is a purely quantum mechanical phenomenon, 
without any classical analogs. 
Yet, the observed longitudinal magnetoresistance is a macroscopic effect, seen in a large sample. Such macroscopic quantum phenomena, well-known examples being 
superconductivity and quantum Hall effect, are typically observed only at very low temperatures (high-temperature superconductivity is a notable exception).
The fact that the chiral anomaly manifestation in Na$_3$Bi is observed at temperatures as high as 90 K makes it especially interesting, and potentially useful technologically. 
 \bibliography{references}

\begin{thebibliography}{10}

\bibitem{Adler69}
S.~L. Adler, {\it Phys. Rev.\/} {\bf 177}, 2426 (1969).

\bibitem{Jackiw69}
J.~S. Bell, R.~Jackiw, {\it Nuovo Cimento A\/} {\bf 60}, 4 (1969).

\bibitem{Nielsen83}
H.~Nielsen, M.~Ninomiya, {\it Physics Letters B\/} {\bf 130}, 389  (1983).

\bibitem{Wan11}
X.~Wan, A.~M. Turner, A.~Vishwanath, S.~Y. Savrasov, {\it Phys. Rev. B\/} {\bf
  83}, 205101 (2011).

\bibitem{Burkov11-1}
A.~A. Burkov, L.~Balents, {\it Phys. Rev. Lett.\/} {\bf 107}, 127205 (2011).

\bibitem{Shen14}
Z.~K. Liu, {\it et~al.\/}, {\it Science\/} {\bf 343}, 864 (2014).

\bibitem{Hasan15-Sci}
S.-Y. Xu, {\it et~al.\/}, {\it Science\/} {\bf 349}, 613 (2015).

\bibitem{Lu15-Sci}
L.~Lu, {\it et~al.\/}, {\it Science\/} {\bf 349}, 622 (2015).

\bibitem{Spivak12}
D.~T. Son, B.~Z. Spivak, {\it Phys. Rev. B\/} {\bf 88}, 104412 (2013).

\bibitem{Burkov15-2}
A.~A. Burkov, {\it Phys. Rev. B\/} {\bf 91}, 245157 (2015).

\end{thebibliography}
 \bibliographystyle{Science}
 
 \end{document}